\RequirePackage{fix-cm}
\documentclass[smallextended]{svjour3}       
\smartqed  
\usepackage{graphicx}
\usepackage{graphicx, color, amsmath, url, ulem}
\usepackage{natbib}
\usepackage[T1]{fontenc}

\spnewtheorem{hyp}{H}{\bf}{\it}

\renewcommand{\PACSname}{\textbf{JEL classification}\enspace}

\begin{document}

\title{Towards more effective consumer steering via network analysis}


\author{Jacopo Arpetti \and Antonio Iovanella}


\institute{Jacopo Arpetti, corresponding author (http://orcid.org/0000-0002-3448-1055),
\\Antonio Iovanella, (http://orcid.org/0000-0001-8147-3747) \at
	Department of Enterprise Engineering, University of Rome Tor Vergata,\\
	Via del Politecnico, 1 - 00133, Rome, Italy\\
	Tel.: +39-06-72597788\\
	\email{(jacopo.arpetti, antonio.iovanella)@uniroma2.it}
}

\date{Received: date / Accepted: date}

\maketitle

\begin{abstract}

Increased data gathering capacity, together with the spread of data analytics techniques, has prompterd an unprecedented concentration of information related to the individuals' preferences in the hands of a few gatekeepers. In the present paper, we show how platforms' performances still appear astonishing in relation to some unexplored data and networks properties, capable to enhance the platforms' capacity to implement steering practices by means of an increased ability to estimate individuals' preferences. To this end, we rely on network science whose analytical tools allow data representations capable of highlighting relationships between subjects and/or items, extracting a great amount of information. We therefore propose a measure called Network Information Patrimony, considering the amount of information available within the system and we look into how platforms could exploit data stemming from connected profiles within a network, with a view to obtaining competitive advantages. Our measure takes into account the quality of the connections among nodes as the one of a hypothetical user in relation to its neighbourhood, detecting how users with a good  neighbourhood -- hence of a superior connections set -- obtain better information. We tested our measures on Amazons' instances, obtaining evidence which confirm the relevance of information extracted from nodes' neighbourhood in order to steer targeted users.

\keywords{Data value \and Network-driven economy \and Steering \and Networks theory \and Nearest neighbour degree}
\PACS{D83 \and D85 \and L11}
\end{abstract}


\section{Introduction}
\label{intro}

Among the revolutions featuring the last century, the most important one, which marked every economic sector, impacting on every aspect of our daily life, is the Web revolution. Nowadays, starting with its 2.0 version, the Web has made individuals no longer mere users of information available on the internet, but rather -- with the advent of internet platforms and mobile apps offering access to a wide array of services such as search engines, maps, and music or video on demand ~\citep{CouncilofEconomicAdvisers-CEA2015} -- true data forgers, often of a significantly personal nature~\citep{jentzsch2017secondary}. In this context, data sharing increases to a global reach~\citep{Acquisti2016}, blurring the distinction between the digital and the physical world, between online and offline.

Thanks to the increased data gathering capacity and the spreading deployment of data analytics techniques, a huge amount of information related to individuals' preferences -- of which no trace was previously left in the real world~\citep{Pagallo2014} -- is today in the hands of a few gatekeepers. This allows them to potentially influence and steer individuals' purchase choices, as well as to charge -- in case they are e-commerce platforms -- different prices to different customers \citep[pp. 445, ][]{Acquisti2016, CouncilofEconomicAdvisers-CEA2015, kshetri2014big}, who are clustered via different techniques. 

As a matter of fact, the more data platforms manage to gather, the more effective their attempt will be to shift from a third price discrimination degree (different prices are charged to different socio-demographic groups) to a perfect discrimination, that is the condition in which it is possible to set different prices for different individuals or to nudge the same consumers towards the purchase of a certain good o service based on their price sensitivity~\citep{shiller2014first, kramer2017digital, regner2017privacy}.
It happen indeed to be addressed by product recommendations based on one's past online purchases, to be targeted by real-time advertisements reflecting past browsing behaviour, or to be the recipient of tailored search results stemming from individual queries~\citep{Levin2011}. 
As a matter of fact, platforms collect data, the more they can move towards perfect discrimination. The proof of the progressively enhanced platforms' capacity to implement discriminatory practices lies with their increased ability to estimate individuals' preferences~\citep{lu2012recommender, lu2015recommender}.

The aim of the present paper is to look into whether, by using network representations and exploiting network properties, platforms could use data acquired from customers to design even more efficient price discrimination practices, or to better nudge and steer consumers towards their purchases. It appears indeed that, by letting information flow throughout the entire network, and thus taking into account the preferences of nodes which are closer to the targeted one (i.e. its {\it alters}), it would be possible for platforms to capture a wider amount of information compared to a situation where individuals are divided into clusters as it currently generally happens. As a matter of fact, considering the whole network and not just its subsets (thought clustering techniques, thereby items and/or subjects are divided into clusters, being considered as a portion of the entire universe) it is reasonable to think that the correspondingly greater amount of information, collected in the entire universe, could lead to better profiling results.

The theoretical proposal offered in the present paper is mainly based on the concept of {\it network value}\footnote{In the light of the existing literature, the value of a network can be inferred from to Metcalfe's Law~\citep{gilder1993metcalfe} which defines the network value as proportional to the square of its size. Some other laws have been proposed: Sarnoff's and the Reed's laws~\citep{reed1999sneaky} and the Odlyzko's one~\citep{BOT2006}. However, despite its simplicity and some of its limits~\citep{swann2002functional, BOT2006}, Metcalfe's law remains a reliable tool~\citep{madureira2013empirical, van2016testing}, which has been used, for example, in order to figure out Facebook's network value~\citep{metcalfe2013metcalfe, zhang2015tencent}.}, which has been used in order to estimate the information available within the system, as well as the way how platforms exploit data stemming from connected profiles in a network.

The stemming of information availability from the network connection patterns, is taken into consideration starting from the observation that a single node is not only affected by its alters (i.e. its neighbours) but also by the alters of its alters~\citep{cerqueti2018new}. We therefore make use of some related network measures, and in so doing we provide several brand-new measures, shifting from the concept of value to the concept of information patrimony, both at the node and network levels.

The aim of the present paper is not to provide a methodology aimed to set different prices for different users, but rather to shed some light on how network systems could potentially provide a greater amount of information than expected.

The paper is organised as follows: Section~\ref{recsystems} offers an overview of the evolution of profiling techniques. Section~\ref{networkprop} provides a review of network properties and their impact on profiling techniques. Section~\ref{theory} is devoted to the outline of certain relevant notations about network theory. Section~\ref{NIP} introduces and discusses our proposal and the related interpretation. Section~\ref{example}  shows some explanatory simulations. The final section offers some conclusive remarks and proposes directions for future research.

\section{The evolution of clustering-based profiling techniques and related limitations}\label{recsystems}

Since the mid-90s e-commerce platforms begun exploiting input provided by customers in order to generate a list of items to be recommended to them~\citep{Kamishima2011}. A technical way through which platforms improve their own algorithms in order to identify individual preferences is represented by {\it Collaborative Filtering} techniques ($CF$)\footnote{A collaborative filtering algorithm implements a recommender system~\citep{lu2012recommender} in order to offer goods or services to a user, starting from the analysis of his/her behaviours (and those of similar users), ending up with advising what the same consumer might find useful, on the basis of the expressed preferences~\citep{Resnick1997, sarwar2001item, Breese2013}, of the choices made by similar consumers, which allows to cluster them~\citep{xue2005scalable}.
Relying on similar form of users' profiling techniques (such as Collaborative Filtering methods) may allow indeed an efficient matching of people and relevant purchase opportunities,~\citep{Levin2011}, although it could also bring about some distortions and disequilibria, up to market failures~\citep{Gertz2002}. For a different take on the functioning of digital markets, see~\citet{Fuller}.}. The aim of platforms is indeed to determine the preferences of individuals not only on the basis of data produced by some targeted ones, but also by using information derived by users related to him by the same purchase choices and preferences. In this regard, already back in 2010, in a special issue of The Economist, reference was made to the fact that Amazon and Netflix were using a statistical technique called {\it collaborative filtering} in order to \guillemotleft make recommendations to users based on what other users like\guillemotright~\citep{TheEconomist2010}. Algorithms aggregate items from similar customers, then eliminate items that a user has already purchased, recommending the remaining items to the same subject\footnote{Amazon's profiling techniques -- their so called item-to-item collaborative filtering -- focus is though not on grouping similar customers on the basis of their individually analysed behaviours, but rather on clustering them via the correlation of similarities between the items that they have chosen: on the basis of a user's purchased and/or rated items, the algorithm attempts to find out similarities with items not chosen yet, then it aggregate them in order to come up with purchase recommendations~\citep{Linden2003}.}.

By profiling consumers through their digital footprints\footnote{Collaborative filtering offers consumption prediction based on the customers' purchase history~\citep{Linden2003}.}, firms can predict individuals' demand as well as their price point sensitivity, ultimately altering the balance of power in their price and value negotiations~\citep{Gertz2002}. Furthermore, sorting out the recommendation puzzle -- deciding what to suggest to whom -- can allow platforms to implement incisive forms of price discrimination\footnote{In this respect, it is worth recalling that the economic literature identifies three types of price discrimination: first-degree price discrimination, which occurs when a seller is a position to charge different prices for each buyer (personalised pricing); second-degree price discrimination, occurring when a consumer self-selects himself by choosing a specific package -- whose per-unit price is dependent on the amount purchased -- that best fits his needs (think about the Netflix package that each individual is free to choose according to his specific needs); third-degree price discrimination arising when sellers propose different prices to different socio-demographic groups~\citep{Cabral2000, acquisti2008identity, Arpetti2018}.}.

One of the most effective ways to find out which item should be offered in relation to the user's needs is grounded on algorithms trained to group or classify together similar objects or similar individuals via collaborative filtering. The goal of algorithms used by platforms, is essentially to lead the decisions made by users, suggesting them a certain good instead of another\footnote{Such goods could be perfect substitutes and an algorithm could decide to show only a part of them, proposing to the consumer the only goods that reflects the {\it Willingness To Pay} (WTP) identified for that profiled user.} in relation to their price sensitivity.

Clustering and classification algorithms (such as those used by CF, like $k$-Means Clustering and $k$-Nearest Neighbours algorithm, $k$-NN\footnote{The two most well-known classification and clustering algorithms -- used by CF -- are, respectively, $k$-nearest neighbour algorithm ($k$-nn) (not to be confused with the ``Nearest Neighbour Degree'', $k_{nn}$, which origins from the network science field and which we will use later in this paper) and $k$-mean clustering~\citep{sarwar2001item, mobasher2001improving}. While the $k$-nn algorithm proposes a classification measuring the distance between similar items or users in order to suggest which of them are closer (in terms of the already purchased item or in terms of users brought together by similar purchase histories) from the targeted user, $k$-mean algorithms are designed to cluster items (and so individuals on the basis of their choices and purchase histories when they are ``user-item''), suggesting to the targeted subject those items surrounding those already purchased by him/her, thus defining the center of the temporary cluster and suggesting the closest nodes~\citep{paterek2007improving, katarya2016collaborative}.}) show though several limitations. As a matter of fact, even if their computational complexity is known to be polynomial in the size of the number of points, such algorithms suffer from major unsolved problems~\citep{xu2005survey, firdaus2015survey}. These latter are mainly related to sparsity of data, high redundancy, inherent noise, sensitivity to the outliers, clusters dimensional heterogeneity; furthermore, there is no metric to assess the quality of the results~\citep{sarwar2001item, nguyen2007improving}.
In addition to the aforementioned scalabilty and spartisy,  recommendation system suffer from the so-called ``cold-start" problem~\citep{nguyen2007improving, castillejo2012social, konstas2009social, lam2010system}. 

Although some authors have tried to sort out such matters via combining data form social network websites (ie information gathered via their explicit social network) and recommendation systems~\citep{konstas2009social, liu2010use}, the aforementioned problems still remain unsolved~\citep{castillejo2012social}. 

\section{Network properties applied to profiling tecniques}\label{networkprop}

As shown by Google search engine, networks' properties might be crucial in resolving recommendation systems limitations~\citep{page1999pagerank, castillejo2012social}. Indeed, platforms may rely on an alternative data representation -- that is, the network representation of the relationships between nodes representing items offered by the platforms and/or users -- based on the {\it network science} (NS) paradigm~\citep{barabasi2013network}. As matter of fact, a formal network representation (i.e. a mathematical representation) allows to organise users or objects as connected via links that make explicit predefined relationships. It is therefore possible to use network science tools and, in particular, those inherited from {\it Social Network Analysis\footnote{Is it worth to specify that the term {\it social network} has a double meaning: it may refer both to a social structure where actors interact networked by means of a web platform and to the scientific discipline of measuring actor interaction on a network as a mathematical object.}}~\citep[e.g.][]{scott2011sage} to build a scheme describing preferences' patterns featuring connections between items and/or users. According to the literature, such representation has become the preferential tool used to study complex systems~\citep{wang2003complex}.

The use of Network Science may improve the quality of recommendations made to consumers via the combination of metadata stemming from the users' activities -- hence referred to the targeted node -- with those gathered from the nodes' neighbourhood and by exploit via networks' properties (see paragraph 4). 

In this regard, we consider that an underlying network expressing the item-user relationship is always potentially available to platforms, since these latter  can rely on two types of networks: an explicit one (used by Facebook and any other social networks whereby network are built and inter-connected) and an implicit one, where the established connections among users are build on their interests or, in general, on their online behaviour~\citep{zhou2014social}.

Thanks to network science it is indeed possible to understand the relations between items\footnote{In order, of course, to suggest them to the platform's users.}, not by measuring the distance ($k$-NN) between individuals or goods and/or services within a cluster, but by considering the totality of connections between nodes (which may represent both items and users) in the entire network (e.g. see~\citet{birke2013social} and \citet{lu2015recommender} and references therein). The user-user CF method, for example, implies that it is possible profile individuals by inferring the data they ceded so to put them in a group of subjects with similar characteristics, in order to proceed to $k$-NN measurements and find out what the nearest neighbours are. In the case of CF, clusters are created in order to reflect the platform's needs on the background of the users' (or items') characteristics: applied to consumption forecasts, both behaviour predictions and item suggestions are ``bent'' by the use of Big Data, which allow the platform to modify the structure of the network through a continuous remodelling of the relevant clusters. Conversely, NS techniques allow the network structure to remain rigid, while data flow in a unique cluster (the entire network) and information stems from the measurement of connections in an immutable structure.  

Network analysis and its properties -- along with the data analysis made by any other recommendations system, although with a different computational effort -- can be used by platforms in order to manage some of the factors influencing the purchase decisions by individuals plunged in the internet context\footnote{i.e. behavioural biases to which consumers are generally subject.}, as well as a range of other elements that come into play as regards such choices. In such sense, due to the high level of information asymmetry~\citep{Akerlof1970, tsai2011effect, mavlanova2012signaling} associated to the deep consumers' unawareness of how digital footprints remain on the web\footnote{i.e. purchase and internet history, GPS data stemmed from mobile devices, etc.}, networks properties can be used to perform profiling and grouping practices, to the point that a seller could decide to offer a certain item instead of another, hiding the existence of perfect substitute goods, nudging individuals to the purchase of a more expensive but equivalent item on the basis of a connections analysis (hence an analysis of the connections between the goods purchased by the individual and those connected to them) aimed at estimating the users' price sensitivity~\citep{mikians2012detecting, mattioli2012orbitz}\footnote{In this regard, the UK Competition and Market Authority stated that: \guillemotleft Firms could also seek to discriminate between customers using competitive variables other than price. [...] The collection of consumer data may enable firms to make judgements about the lowest level of quality needed by consumers/groups of similar consumers. This may enable a firm to engage in quality discrimination where quality differences are not reflected in the prices of goods or services. Firms may do this by restricting the products that are displayed to consumers or by varying the order in which products are listed on their website to display relatively poorer or better quality products first depending on the information they collect about consumers\guillemotright~\citep[][pp. 93]{CMA-CompetitionandMarketsAuthority2015}.}.

When asked how e-commerce websites could implement price ``personalisation'' in relation to each consumer so to squeezing their surplus, ~\citet[][pp. 306]{Hannak2014} answered that two practices had been conceived by those players: \guillemotleft price discrimination -- customising prices for some users -- [...] (and) price steering -- changing the order of search results to highlight specific products\guillemotright.

Network science may ensure a more effective implementation of these practices because evaluations made by the platforms on what item to show or not, would be carried out in relation to the whole universe of items available to all users and not only to those addressed -- at a given time -- to a certain group of individuals. In this context it is therefore appropriate investigating how the use of NS could lead to an evolution of profiling practices grounded on the collection of habits and preferences, in the light the unprecedented capacity of online platforms to collect digital footprints left by individuals~\citep{fuller2018privacy, acquisti2008identity}, even considering constraints such as clustering limitation. In such sense, due to the presence of some constraints, it is improper to assert -- contrary to what claimed by a wide part of the literature (Council of Economic Advisers - CEA, 2015) -- that we are assisting to a shift from third-degree price discrimination to totally personalised prices just because of the increased availability of behavioural data. There are indeed at least three obstacles -- in addition to the fact that platforms do not implement similar strategies due to reputation motivation as reported in~\citet{CouncilofEconomicAdvisers-CEA2015} -- standing between perfect price discrimination and the forms of discrimination that platforms currently implement~\citep[][pp. 96-99]{Ezrachi2016}: Lack of data\footnote{A platform should be able to estimate each consumers' reservation price, variable that is not directly observable but just inferred from. Moreover, not all platforms avail of the means to constantly observe every move made online by each consumer, while none of them can estimate their WTP for each item. As stated by Ezrachi and Stucke in ``Virtual Competition":\guillemotleft One impediment to perfect discrimination is insufficient data. Although the algorithm has a lot more personal data that brick-and-mortar retailer of twenty years ago, the algorithm still has insufficient data for any particular customer: the customer may never have bought the item before; and the customer's behavior may never have signaled how much or she is willing to spend to accurately predict an individual's reservation price would require sufficient data to identify and measure each of many variables that affect the reservation price\guillemotright. ~\citep[][pp. 96-97]{Ezrachi2016}.}; Irrationality\footnote{In order to implement a perfect price discrimination practice, it would be necessary to predict how individuals are subjected to biases and heuristics, that can change their preferences and affect their reservation prices: an algorithm whose predictions are based on preexisting reservation prices cannot ignore how such elements could change~\citep{arrow1958utilities, kahneman1986rational, herbert1955behavioral, simon1990bounded}.}; Sufficient sample size\footnote{An algorithm must have an adequate amount of data so that a perfect price discrimination practice can be performed. To this purpose, it is presumable that algorithms avail of the necessary data sources to infer information, hence to make predictions on individuals' purchase preferences related to products routinely bought. However, as pointed out by Ezrachi and Stucke, the platforms do not have sufficient data about scattered purchases: when a purchase is not cyclically performed, it is more complex to elaborate forecasts on individual behaviour because of the absence of a sufficient amount of \guillemotleft trial-and-error\guillemotright data up to the point to detect every variable needed to figure up the reservation price of each individual for a corresponding good (think, for instance, about the rate at which individuals buy a PC screen~\citep[][pp. 99]{Ezrachi2016}).}. In the light of such obstacles, platforms must put in place practical solutions so that preferences can be detected and reserve prices -- which are not directly deducible -- approximated~\citep{Shiller2015}~\citep[][pp. 96]{Ezrachi2016}\footnote{In order for this to happen, segmentation strategies and individuals' clustering practices are put in place to include each individual in a specific group of consumers who share similar preferences and price sensitivity, assuming that grouping persons with analogous tastes and purchase behaviours in small clusters allows a better approximation of their reservation prices~\citep{Ezrachi2016a}}. Starting from the purchase behaviour patterns charactering each group to which individuals have been allocated to, platforms predict consumers' desires and ``their next moves'', prompting further consumption through the suggestion of specific products or services that have proven to be popular among the choices made by individuals. Such items, identified through the behaviours of consumers clustered in homogenous groups (pooled e.g. by preferences, behavioural patterns, purchase history and price reservations) are selected based on the corresponding utility levels express by individuals or, when items are bundled (or complementary), via the utility level related to one of them, influencing the utility of the other~\citep{Zhao2016}.

What is then the role of NS in the light of the platforms' profiling practices carried out in order to detect users' preferences and to nudge them to consumption?
Although the White House report on Big Data and differential pricing states that steering practices do not make use of information concerning potential buyers at the individual level, and this is the reason why those practices should not be considered any longer as prevailing\footnote{Despite the fact that, the same report further explains how more complex it is to infer information about Internet users while having access to their data (such as the user's IP address, or the kind of operating system) but without being able to define and their willingness to pay.}~\citep{CouncilofEconomicAdvisers-CEA2015}, that does not mean that steering practices\footnote{To be understood as the possibility to decide what to show to individuals and guiding their choices in this sense.} are not still used (see~\citet{mikians2012detecting}). 

Although the primary goal of the present contribution is not the exploration of how the increased observation of the individuals' purchase choices and expressed preferences (hence the greater amount of data which platforms possess on them) could lead to more performing steering, nudging and price discrimination practices, we show how social network science has became crucial in predicting consumers consumption patterns in the light of  networks features which can be better understood through some measures that we propose below.

\section{Methodological prerequisites}\label{theory}

The next sections are devoted to the conceptualisation of our proposal and we therefore provide further below some preliminary notations, for the convenience of the reader. 

\begin{figure}
\centering
 \includegraphics[scale = 0.8, trim={10cm 20.5cm 14cm 4cm}]{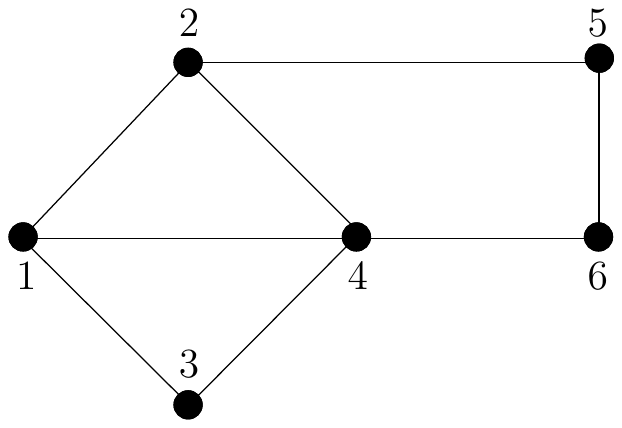}	
\caption{An example of a basic graph with nodes $n= 6$ and links $m = 8$.}
\label{fig:1}
\end{figure}

The classical mathematical abstraction of a network is a graph $G = (V,E)$, where $V$ is the set of $n$ nodes (or vertices) and $E$ is the set of $m$ links (or edges) outlining the relationships among the nodes. For instance, taking into account the network of users in an online marketplace, nodes will be made up of individuals linked among them by the purchased items. Conversely, should, for the same online marketplace, be considered the network of items, nodes will represent goods or services linked together because bundled or, in an alternative setting, purchased by the same user. In this regard, Figure~\ref{fig:1} provides a basic example of a users' graph relating to a generic online marketplace with $6$ nodes corresponding to as many individuals, linked among them by $8$ identical purchase choices (namely  the purchase of the same goods or services) represented by as many links. In such setting, information spans from node to node, within the network, via the links' pattern~\citep{bakshy2012role, galati2019framework}.

It is worth mentioning that, according to the literature, both the word ``network''  and  ``graph'' can be used interchangeably. While the first is generally associated to real systems, such as the WWW or networks of individuals, ``graph'' is mainly used to analyse the mathematical representation of a network~\citep[pag. 45,][]{barabasi2016network}.

We refer to a node by an index $i$, meaning that we allow a one-to-one correspondence between an index in $\{1, \ldots, N\}$ and a node in $V$. The set $E$ can be conceptualised through the adjacency matrix $A = (a_{ij})_{i, j=1, \ldots, N}$, whose generic element $a_{ij}$ is equivalent to $1$ -- if the link between $i$ and $j$ exists -- or, otherwise, to $0$.

In this paper, we examine undirected and unweighted networks, thus $a_{ij} = a_{ji}$, for each $i, j = 1, \ldots, N$.  In other words, considering the previous graph in Figure~\ref{fig:1}, the index represents the identification number of a generic user, while the adjacency matrix is a table used to gather the relations between all purchase choices expressed by users in a given moment, identifying a formal relationships between them.

In graphs, a usual measure is given by the number of relationships owned by a user. In the case of Figure~\ref{fig:1}, user $1$ has three relationships, namely with users  $2, 3$ and $4$. For a generic user $i$, such relations are expressed as a measure called {\it degree}, with symbol $d_i$, while the nodes linked to node $i$ are called the {\it neighbour} of $i$ (in symbols, $N(i)$). It is clear, that $d_i$ is an integer being zero when the user has no established relations.

All the degrees can be gathered in a vector, whose elements are ordered in a non-decreasing mode. This vector, called {\it degree sequence} with symbol $D_G$, is -- in our example -- $D_G = \{4, 3, 3, 2, 2, 2\}$. On the basis of such sequence, it is possible to compute the traditional average value $\langle d\rangle$, as well as the average of the squared values $\langle d^2\rangle$, that are -- in our example -- respectively, $\langle d\rangle =  2.67,$ and $\langle d^2\rangle = 7.67$.

In order to study networked systems properties, it is possible to exploit a method allowing the creation of a random network by using a given degree sequence. To this end, one of the most common and used tool is the {\it configuration model}~\citep{newman2018networks}. This latter, defined as a generalised random network model whit a given degree sequence, is build up by means of a simple algorithm~\citep[e.g. pag. 139,][]{barabasi2016network}.

Configuration model can also be generated considering degree sequences stemming from a given distribution. Thanks to such feature, it is possible to generate specific networks taking into account,  e.g, Poisson's or Power-Law distributions, which are the widely used degree configurations in network science.

The purpose of the configuration model is that of leveraging some of its known properties in order to obtain general results suitable for building a new theoretical framework\footnote{The aim of the present paper it is not to analyse the above mentioned algorithm~\cite[e.g. see][]{hakimi1962realizability} neither to investigate the mathematical properties allowing the graph creation from a given sequence~\citep{EG1960}.}.

Considering the configuration model properties, we therefore suppose that a platform may generate its own network (thus, the corresponding graph) starting from a database containing a given number of users (therefore, the length of the degree sequence) and their relative purchase choices (hence, the elements within the degree sequence).

In the light of such elements, we suppose that an enterprise, and in particular a platform, can always build up its own network considering consumers' data as well as data related to items offered to them. In this regard, a network can be: (i) explicitly available, when the enterprise delivers multi-users products or services (e.g., Facebook, LinkedIn, Twitter and MySpace); (ii) not explicitly available, since the firm delivers single user products or services (e.g. Amazon, Last.FM, Outbrain and Color)\footnote{In such case the network can be generated exploiting analytical methodologies~\citep{zhou2014social}.}.

Among the properties offered by the configuration model, we used -- for our purposes -- that of the {\it average degree of a neighbour}, represented with symbol $k_{nn}$. For a generic node of the network, it holds that~\citep{catanzaro2005generation}:

\begin{equation}
\label{neighbor}
k_{nn} = \frac{\langle d^2\rangle}{\langle d\rangle}
\end{equation}

The average degree of a neighbour is widely used to study dependencies between neighbours nodes' degrees in a network, affirming that, considering a node $i$, $k_{nn}$ is independent from the node's degree but exclusively dependent on the global network's characteristics $\langle d\rangle$ and $\langle d^2\rangle$. Besides being easily computable, such measures provide an overall view of the network, making $k_{nn}$ a simple and powerful global measure of it.

Generally, $k_{nn}$ turns out to be greater than the average degree of a node. As a matter fact, considering ${\langle d^2\rangle}/{\langle d\rangle} - {\langle d\rangle} = ({\langle d^2\rangle} - {\langle d\rangle}^2)/{\langle d\rangle} = \sigma^2/{\langle d\rangle} > 0$, since: (i) the variance is non-negative unless the network shows the same degree for all its nodes; (ii) $\langle d\rangle$ is greater than zero unless all the nodes have zero degree. This holds that ${\langle d^2\rangle}/{\langle d\rangle} > {\langle d\rangle}$, showing a result known as the {\it friendship paradox}, according to which \guillemotleft your friends have more friends than you do\guillemotright~\citep{feld1991your}. Such bias is due to an over-representation of high-degree nodes in comparison with low-degree ones during the calculation.

Please note that Formula (\ref{neighbor}) refers to the general case of uncorrelated degree sequences~\citep{barabasi2016network, newman2018networks}. Networks generally show a degree-degree correlation  which can display a positive degree correlation in the case of social networks or a negative one, for instance, in the case of networks related to technological and innovation research projects. In such context, the average degree of a neighbour of node $i$ is computed as~\citep{pastor2001dynamical}:

\begin{equation}\label{knni}
k_{nn,i} = \frac{1}{d_i}\sum_{j}d_ja_{ij}
\end{equation}

A further measure can be expressed considering sets of nodes for identical degree. In this case, the average degree of a neighbour for nodes with degree equal to $k$, is defined as:

\begin{equation}\label{knnd}
k_{nn}(k) = \frac{\sum_{i:d_i= d}k_{nn,i} }{\sum_{i:d_i= d}1}
\end{equation}

Considering the example shown in Figure~\ref{fig:1}, Equation~\ref{neighbor} brings to $k_{nn} = 2.9$, Equation~\ref{knni} brings to $k_{nn, 1} = 3.0$, $k_{nn, 2} = 3.0$, $k_{nn, 3} = 3.5$, $k_{nn, 4} = 2.5$, $k_{nn, 5} = 2.5$ and $k_{nn, 6} = 3.0$, while Equation~\ref{knnd} to $k_{nn}(2) = 3.0$ for the set of nodes of degree 2, $k_{nn}(3) = 3.0$ for the set of nodes of degree 3 and $k_{nn}(4) = 2.5$ for the set of nodes of degree 4.

$k_{nn}(k)$ nature highlights two distinctive structural network properties with reference to the presence of degree correlation. If $k_{nn}(k)$ is an increasing function of $k$, such setting identifies {\it assortative mixing}, showing that high-degree nodes are preferentially connected to high-degree nodes, while the low ones to low ones. On the other hand, a decreasing trend of $k_{nn}(k)$ identifies {\it disassortative mixing} showing nodes with high connections having neighbours with lower ones.

\section{Neighbours' Information patrimony}\label{NIP}
\label{sec:cms}

In this section, we introduce our definition of Neighbours' Information Patrimony ($NIP$).

To this purpose, we start from the concept of ``network value'' which has been mostly defined on the basis of Metcalfe's law. Comparing the weight of a single node with the entire network, our definition allows the estimation of such node's ``network share'' by using information related to those nodes to which it is connected. Such approach intends to build on the fact that, generally, each node possesses data~\citep[also called {\it metadata}, e.g. see][]{Peele1602548} through which the enterprise can infer collective information about clients (e.g. geographical localisations, preferences, gender, etc.) to be used, for instance, to implement price discrimination practices. 

In this regard, we propose to consider the network's neighbourhood of a node as a proxy of the similarities among clients, as network clustering techniques do~\citep{cerqueti2018new}. In other words, links between nodes attest a certain level of preferences and/or similar tastes, thus making it possible to extend the chain of relationship from the node's neighbours to the ``neighbours of its neighbours'' (the ``alters of the alters''), shifting the network share of a node from its singularity to its linked environment and, then, to the whole collective patterns of interactions.

Our model builds on the assumption that, if an enterprise intends to gather information about users with similar preferences, it may avoid burdensome techniques (such as cluster analysis or density algorithm), thanks to network properties. In the light of such properties, also knowing a single node-related information, it is possible to acquire data on the neighbour nodes, hence about users with similar characteristics.

We start from the Metcalfe's law which affirms that the value $\mathcal{V}$ of a network $G$ is proportional to the square of the nodes' number, i.e. $\mathcal{V} \propto n^2$~\citep{gilder1993metcalfe}. Stemming from the ethernet network context, the Metcalfe's law definition, considers all nodes as mutually linked\footnote{Therefore, $n$ nodes are so connected to other $n-1$ nodes, thus the value is proportional to $n^2$}.  In general terms, networks are not densely connected, hence estimating the value of a network as $n(n - 1)$ seems to be unrealistic. 

In our setting, considering how available information to the enterprise (or the platform) is given by the amounts of connections set present in a network, it is reasonable to think that all users are not mutually connected, nor are items, or users' preferences in terms of goods or services. By means of the well known ``handshaking lemma''~\citep[e.g.][p. 4]{bollobas2013modern}, we obtain that the sum of all network's degrees is equal to twice the numbers of links, i.e. $\sum_{i \in V} d_i = 2m$. With this setting, we can assume that the object shared in the network is the information patrimony $IP$, thus gathering all the comments so far we obtain the value of $IP$ for the generic node $i$ as:

\begin{equation}\label{nsi}
IP_i = \frac{d_i}{2m}
\end{equation}

It is worth noting that we are reducing the market extension to its real dimension and in case of a complete network $m = n(n-1)/2$, thus the  
denominator of Equation (\ref{nsi}) becomes again proportional to $n^2$, as postulated by the original definition of Metcalfe's law.

Summarising, we are tailoring Metcalfe Law's with a view to estimating the current value of a network, such as $\mathcal{V} \propto 2m$ and the value $\mathcal{V} \propto n^2$ remain still valid for a complete network.

%
%
%


\subsection{Uncorrelated networks}

It is now possibile to introduce the collective effects of alters and we therefore define the {\it Neighbours' Information Patrimony} $NIP_i$ of node $i$ as the sum of its information patrimony with the information patrimonies of all its alters, i.e. the nodes $j$ in its neighbourhood $N(i)$. In formula:

\begin{equation}
\label{CMS1}
NIP_i = IP_i + \sum_{j \in N(i)}IP_j = \frac{d_i}{2m} + \sum_{j \in N(i)}IP_j = \frac{d_i}{2m} + \frac{1}{2m}\sum_{j \in N(i)}d_j
\end{equation}

In Equation (\ref{CMS1}) we approximate the information patrimony of $N(i)$ as $\sum_{j \in N(i)}d_j \simeq d_i \cdot k_{nn}$, that is $d_i$ time the average degree of a neighbour:  

\begin{equation}
NIP_i \simeq \frac{d_i}{2m} + \frac{d_i}{2m} k_{nn} = \frac{d_i}{2m}(1 + k_{nn})
\end{equation}

Finally, using Equation (\ref{neighbor}) for $k_{nn}$ we obtain:

\begin{equation}
\label{CMS}
NIP_i = \frac{d_i}{2m}\bigg(1 +  \frac{\langle d^2\rangle}{\langle d\rangle}\bigg) 
\end{equation}

In Equation (\ref{CMS}) we state that the neighbour information patrimony of a node $i$ is given by the value of its information patrimony, multiplied for a value which captures the level of connectedness given as an overall measure of the whole network. This represents a peculiar measure concerning the entire network $N$, defining what the information is worth within it. We call such measure $NIP_N$:

\begin{equation}
\label{netv}
NIP_N = \bigg(1 +  \frac{\langle d^2\rangle}{\langle d\rangle}\bigg) 
\end{equation}

Since ${\langle d^2\rangle}/{\langle d\rangle} > 0$ and, excluding network with all nodes with the same degree, we have that $NIP_N > 1$.

Please also note that in a complete network -- where all nodes are connected to one another -- the number of links equals $n(n-1)/2$, and for every node we obtain that $IP_i = 1/n$, while $NIP_i = 1$. Such results mean that if each node of a network is mutually connected to the others, then it possesses the whole network information patrimony\footnote{In fact, in this case each node has degree $d_i = n -1$, then $IP_i = d_i/2m = (n-1)/n(n-1) = 1/n$, while $\langle d^2 \rangle = (n-1)^2$ and $\langle d \rangle = n-1$ thus $NIP_i = (1/n) (1 + (n-1)^2/(n-1))= 1$.}.

\subsection{Correlated networks}

As a general case, taking into consideration equation (\ref{CMS1}) we approximate the network sharing of $N(i)$ as $\sum_{j \in N(i)}d_j \simeq d_i \cdot k_{nn, i}$, that is equal to $d_i$ time the average degree of a neighbour of node $i$. Thus, the {\it Network Information Patrimony} is:

\begin{equation}\label{cnei}
NIP_i \simeq \frac{d_i}{2m} + \frac{d_i}{2m} k_{nn} = \frac{d_i}{2m}(1 + k_{nn, i})
\end{equation}

When we move to sets of nodes for identical degree, we have to consider Equation (\ref{knnd}) instead of Equation (\ref{neighbor}) and Equation (\ref{CMS1}) which becomes:

\begin{equation}
\label{cne1}
NIP_i(d) \simeq \frac{d_i}{2m} + \frac{d_i}{2m} k_{nn}(d_i) = \frac{d_i}{2m}\bigg(1 + k_{nn}(d_i)\bigg)
\end{equation}

\subsection{Discussion and implication of the network information patrimony}

In the previous section we introduced three new measures for a given network; one referred to uncorrelated ones (Eq.~\ref{CMS}) and two to correlated networks (Eq.~\ref{cnei} and Eq.~\ref{cne1}). In both cases, we started from the concept of the market share of a node $i$, taking into consideration the average degree of neighbours as a proxy capable to capture the meso-scale connectiveness effects.

Considering a network, the degree correlation is a phenomenon capable to detect the tendency between nodes with similar degree to associate with each other. In a social networks environment, such tendency is called homophily or assortative mixing~\citep{newman2018networks}. A network exhibits assortative mixing if nodes with a high number of connections tend to link with large degree nodes; on the contrary, we are in presence of disassortative mixing when nodes with high connections have neighbours with lower ones. This latter is a frequent feature in many technological networks~\citep{newman2002assortative, d2012robustness}.

Network assortativity is computed through the Pearson correlation coefficient $r$ between two nodes connected by a link with positive $r$ with a value in the range $0 \leq r \leq 1$; while in case of network disassortativity, $r$ is negative with a value between $−1 \leq r < 0$~\citep{newman2003structure}. 

When a network is not correlated\footnote{For instance, when neutral with respect to degree correlation, it does not present assortative mixing ($r = 0$)}, the value of $IP_N$ is just a multiplier for the market share of a node $i$, which is identical for each node. Although such element constitutes a unique measure of a network, the fact that it highlights the connectiveness' degree is of little interest, since we are considering the centrality degree\footnote{The centralisation degree of a node is merely its degree, i.e. the number of links connected to it. In Social Network Analysis literature it is emphasised as a measure since is meaningful in many circumstances. For an extensive discussion on this measure and many other centrality measures please refer to~\citet{scott2011sage} or~\citet{newman2018networks}.} where each value is multiplied by a constant.

As soon as a network exhibits a correlation degree, both $NIP_i$ and $NIP_i(d)$ become meaningful measures. While the former captures the relationship between the degree of node $i$ and the average degree of its neighbours ($NIP_i$ is therefore a measure which magnifies -- or shrinks -- the information patrimony of a node $i$ on the basis of its neighbours' quality), the latter considers the average of all nodes' neighbours degree $d$ (it therefore measures the network effect of those nodes with a degree $d$). $NIP_i$ should be applied in order to detect network's effect of node $i$, while $NIP_i(d)$ would capture the aggregated network effect of the $d$ degree nodes.

Since $NIP_i(d)$ is the average value of all $NIP_i$ of $d$ degree nodes, it could be of interest to explore the scattering around such average value. In such sense, a node with a $NIP_i$ below the average suggests that its neighbours' "value" (to be considered as the amount of information that can be mined from its neighborhood) has a limited effect on its network effect. On the contrary,   if $NIP_i$ is above the average value, it means that the  considered node benefits from the "value" of the neighbours surrounding it. It is indeed admissible, as we will show in the next section, that a node $i$ with degree $d$ can assume values of $NIP_i$ greater than any nodes of degree $d + 1$ (or, under particular circumstances, even greater).

Such observations lead us to conclude that $NIP_i$ and $NIP_i(d)$ are second-order measures~\citep{cerqueti2018investigating, rotundo2014network}, being strongly affected by the presence of peculiar arrangements in the neighbourhood. This represents the main difference with the classical degree centrality measure, that is a first-order measure because it is independent of the network topology. 

Finally, note that the sum of all the market share $IP_i$ trivially sum to $1$, while the sum of $NIP_i$ or $NIP_i(d)$ brings to a sum greater than $1$ since we are considering the neighbourhood effects on each node.

\section{Examples}\label{example}

We herein propose two brief examples. While the first one is based on a synthetic network, the second one considers some real instances based on the {\it Customers Who Bought This Item Also Bought} feature of the Amazon website~\citep{leskovec2007dynamics}.

\begin{figure}[t]
    \centering
    \begin{minipage}{0.5\textwidth}
        \centering
        \includegraphics[scale = 0.4, trim={2cm 2cm 2cm 2cm}]{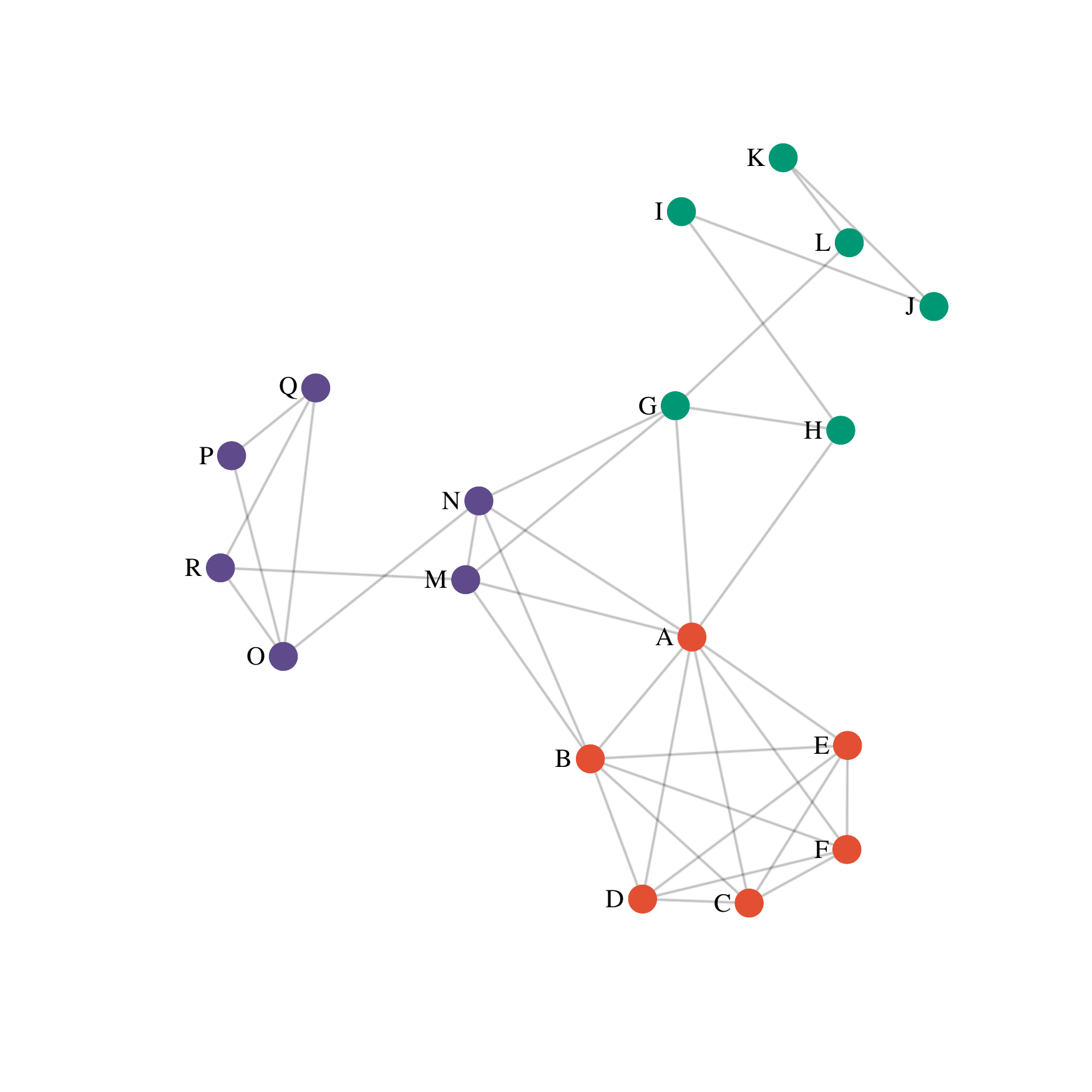}
    \end{minipage}%
    \begin{minipage}{0.5\textwidth}
        \centering
        \includegraphics[scale = 0.33, trim={2cm 2cm 2cm 2cm}]{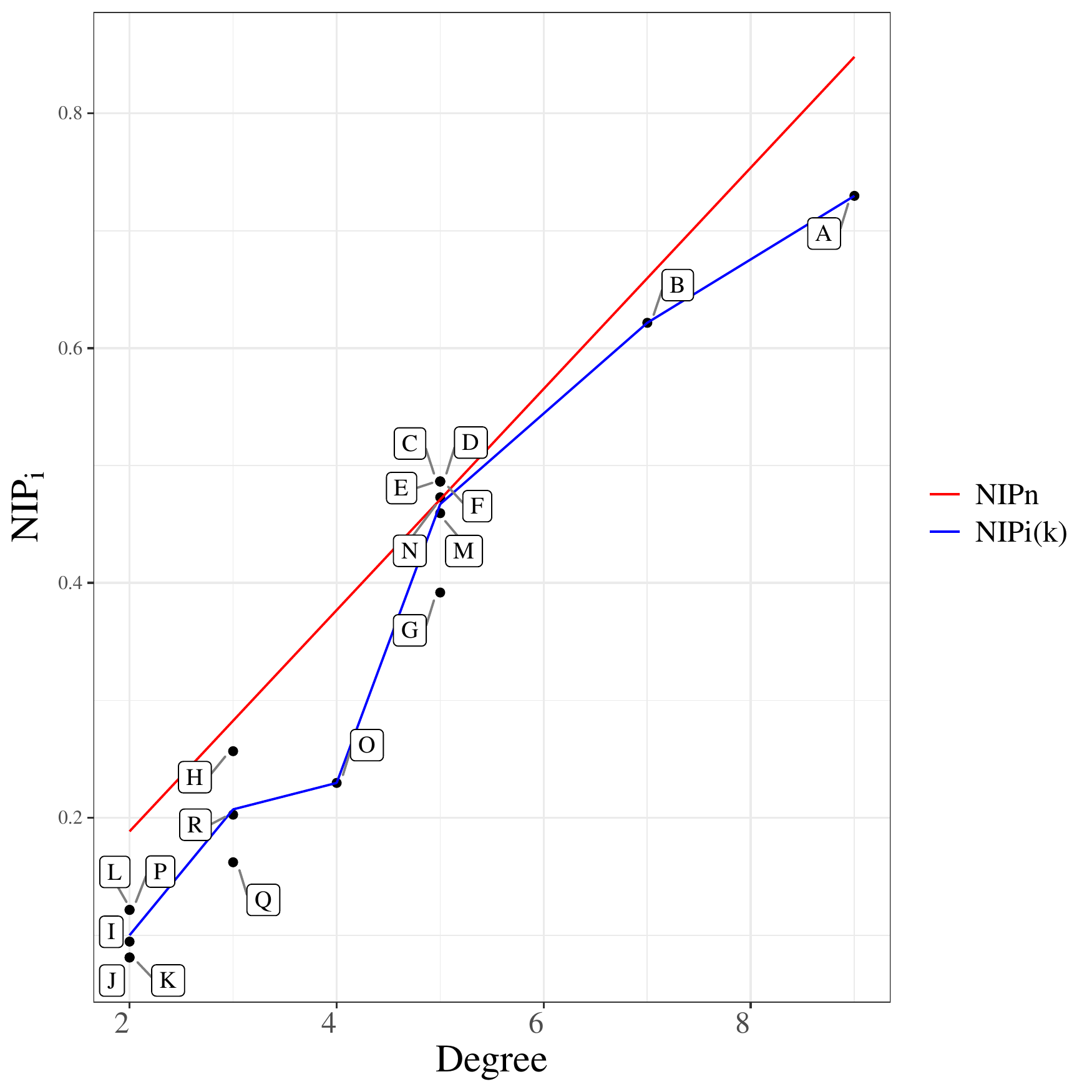}
    \end{minipage}
    \caption{The left figure is an example of a network with three groups of connected users in different combinations: in red a group of mutually connected users, in blue a group with a hub and in green a group wich presents loose connections. The right figure shows the values $NIP_i$ for each node for the graph in the left panel, the values for $NIP_i(d)$ (in blue) and the values for $NIP_i$ considering the network uncorrelated (in red).}
    \label{visualization}
\end{figure}

Let us consider the network as in Figure~\ref{visualization}, which is composed of three groups of nodes (e.g. users or products) linked in order to simulate three different users' agglomeration (e.g. a communities), in turn mutually connected among them through links. With $n= 18$ and $m = 37$, we can easily compute the network information patrimony of nodes by means of Equations~\ref{cnei} and~\ref{cne1}, as shown in the right panel of Figure~\ref{visualization}\footnote{For the sake of completeness, the computational complexity of our measures depend from the calculation of $k_{nn}$, which correspond to a product of a square matrix of size $n$ for a vector of size $n$. It is well known that this calculation is polynomial in the size of the matrix, i.e., it is an $O(n^2)$.}. 

One of the interesting results deducible from the observation of the blue curve in Figure~\ref{visualization}, which is the $NIP_i(d)$,  is that: for a fixed degree, the scattering of the different values of $NIP_i$ highlights nodes under-performing with respect to $NIP_i(d)$; while other over-perform it (e.g., for degree 3, node R is close to $NIP_i(3)$, while node Q is under-performing and H is over-performing). Note also that node H over-performs node O which has degree 4, meaning that the network patrimony derived from the node H neighbourhood has a greater "value" compared to the network patrimony of node O, and this beyond its degree. 

This is a general phenomenon which brings us to formulate the following hypothesis: since nodes with higher network patrimony are connected with nodes with high degree, they benefit from a greater amount of information. In such sense, nodes with $NIP_i$ higher than expected lead us to conclude that they should be basically considered of higher interest for a recommendation system, compared to nodes with $NIP_i$ lower than expected. In other words, $NIP$ permits to discriminate users on the basis of a collective effect based on their neighbourhood.

\begin{figure}[t]
\centering
 \includegraphics[scale = 0.4, trim={0cm 0cm 0cm 0cm}]{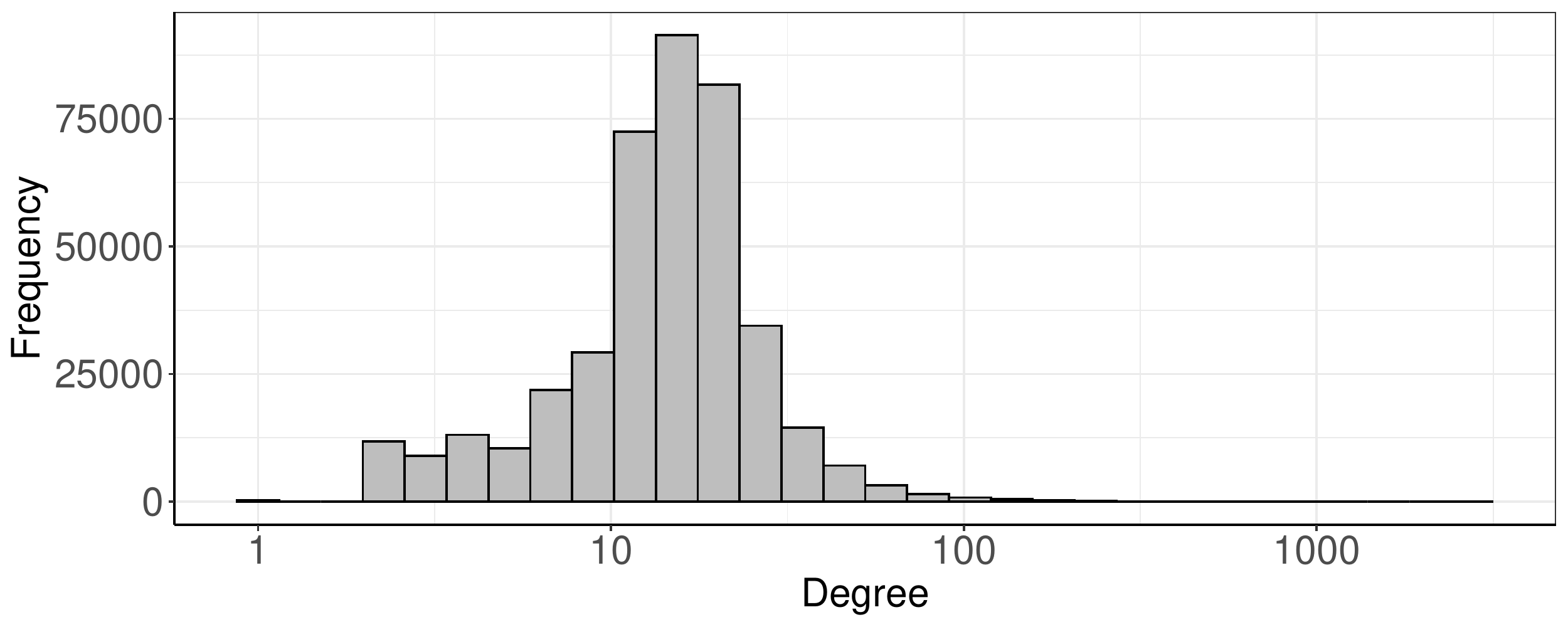}	
\caption{Degree distribution for test instance network {\it amazon0601} (x-axis is in log scale).}
\label{fig:degree}       
\end{figure}

In a second example, we consider the set of instances drawn from the Amazon web site in four different moments in $2003$. Instances are built hypothesising that a product (node) $i$ is co-purchased with product (node) $j$, then the graph contains a link from $i$ to $j$.

Table~\ref{tab:1} summarises basic metrics of such instances\footnote{The data processing, the network analysis and all simulations are conducted using the software R~\citep{team2013r} with the igraph package~\citep{csardi2006igraph}.}, the density\footnote{The density of a network is defined as the number of its actual links divided by the number of potential links which is $n(n-1)/2$. It can assume value from $0$ (empty network) to $1$ (complete network).} and the value for $\langle d\rangle$, $\langle d^2\rangle$, variance, assortativity and $NIP_N$. Since the networks are almost neutral (i.e. $r$ is close to $0$), the calculation of $NIP_N$ is correct and it can be noted how it boosts the magnitude of the information patrimony of each node, even if the network contains very few links, since the density of all instances is of the order of $10^{-5}$.

\begin{table}[t]
\begin{scriptsize}
\caption{Statistics for the Amazon instances}
\label{tab:1}       
\begin{tabular}{lllllllll}
\hline\noalign{\smallskip}
Network         & $n$       &  $m$        & $\delta$ & ${\langle d\rangle}$ & ${\langle d^2\rangle}$ & $\sigma^2$ & $r$ & $NIP_N$\\
\noalign{\smallskip}\hline\noalign{\smallskip}
amazon0302 & 262,111 & 1,234,877 &  $1.79 \cdot 10^{-5}$ & 9.422 & 123.823 & 35.03 & 0.003 & 14.141\\
amazon0312 & 400,727 & 3,200,440 & $1.99 \cdot 10^{-5}$ &15.973 & 505.859 & 250.71 & -0.044 & 32.670\\
amazon0505 & 410,236 &	3,356,824	 & $1.99 \cdot 10^{-5}$ &16.365 & 533.438 & 265.61 & -0.043 & 33.595\\
amazon0601 & 403,394 &	3,387,388 &  $2.08 \cdot 10^{-5}$ &16.794 & 542.731 & 260.68 & -0.043 & 33.317\\
\noalign{\smallskip}\hline
\end{tabular}
\end{scriptsize}
\end{table}

\begin{figure}[t]
\centering
 \includegraphics[scale = 0.4, trim={0cm 3cm 0cm 3cm}]{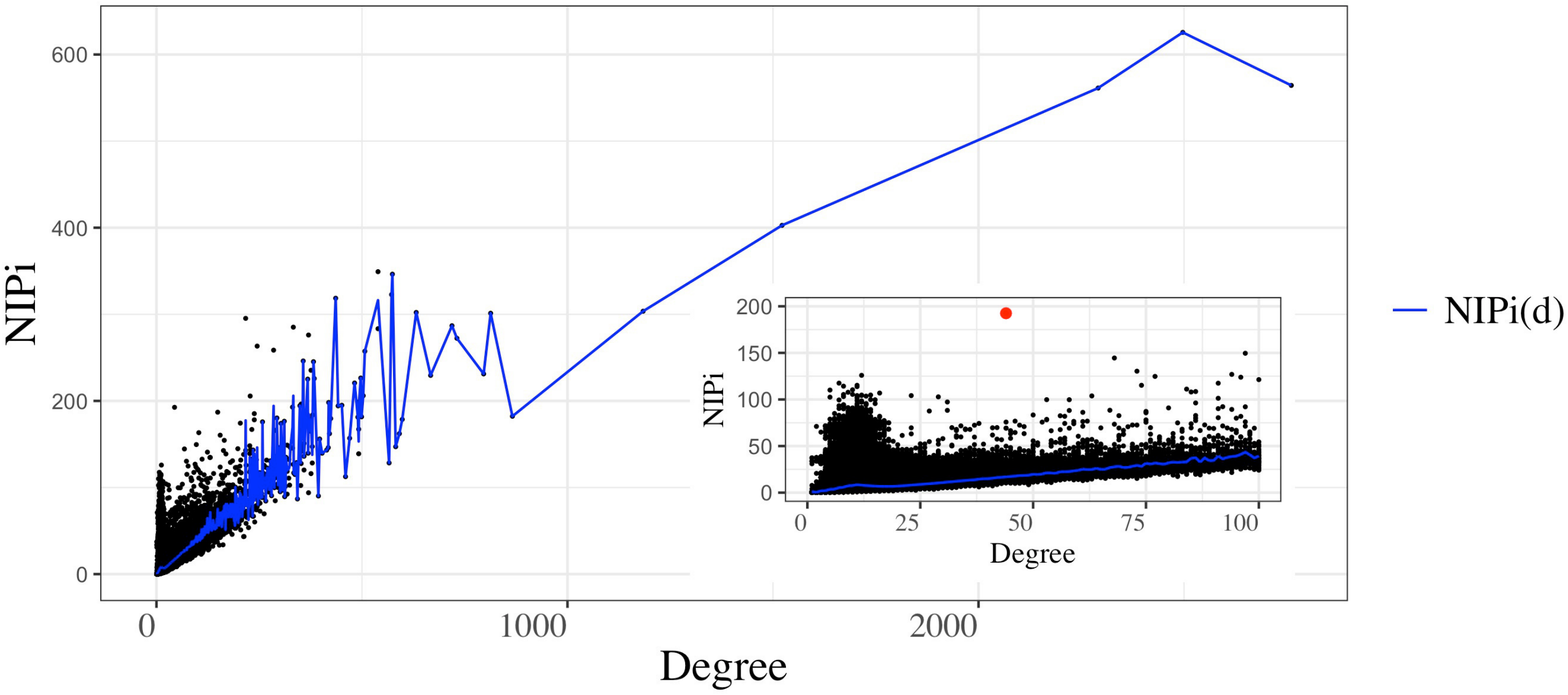}	
\caption{$NIPi$'s value for nodes in {\it amazon0601} network. The inset shows the same values for degree $d_i \leq 100$ and the red point with coordinate $d = 44$ and $NIP_i = 192.61$.}
\label{fig:amaz}       
\end{figure}

Figure~\ref{fig:degree} shows that {\it amazon0601} network instance has a degree distribution shape close to a lognormal, indicating the presence of hub nodes. In Figure~\ref{fig:amaz} we display the distribution of $NIP_i$ and how such values positively correlate with the node degree. Consequently, the majority of nodes with low degree values displays a low network information patrimony. In spite of this, we note a quite strong scattering of the values of $NIP_i$, as shown in the inset of Figure~\ref{fig:amaz}, with some nodes displaying a $NIP_i$ higher than expected, and others displaying a lower one. 

It is also possible to observe that within a class of nodes grouped by the same degree there are some showing different $NIPs$ and -- more surprisingly -- by comparing those classes, it is possible to detect how nodes with a lower degree can have a higher $NIP_i$ than those with a higher degree. In such sense, the red point in the inset in Figure~\ref{fig:amaz} (which corresponds to a node with degree $d = 44$ and $NIP_i = 192.61$) outperforms in terms of network patrimony, showing a $NIP_i$ greater than those of nodes with twice its degree.

\section{Discussions and conclusions}

Building on network science's properties, networks representations can provide significant competitive advantages to those companies which are currently using data without being aware of the gains potentially stemming from an exploitation of such asset.
To explain how companies could benefit from the opportunity to extract much more information than it is reasonable to imagine, we used network science and its formal structure moving from a local vision grounded on clustering and classification analysis, to a global one. Network science properties are indeed useful in order to explain complex systems dynamics, as well as to understand how information flows in a network and how much of it can be extrapolated taking into account a targeted node with respect to its neighbourhood. We therefore exploited networks properties, in order to understand how a platform may use data in order to boost its predictive capabilities in terms of individuals' purchasing capacity and potentially consumers' steering practices.
Such advantage, exploited by platforms relying on a network, can be estimated by measuring the set of their connections as a whole -- that means the links between consumers and/or items within the examined network. In this regard, availing of information on the individual consumer, network platforms can benefit from some of the properties deriving from the network configuration, extracting data on a single node (hence, on a consumer or on an item), but also capturing data stemming from its connections' patterns.

To this end, we proposed an additional measure -- further to the one related to the degree of a node -- taking into account the quality of nodes' connections and consequently of a hypothetical user in relation to its neighbourhood. We therefore exploited a well-known network science concept in order to establish a new analytical quality measure. We measured how users who might enjoy of benefits deriving from a good neighbourhood -- hence of a superior connections set -- may also obtain better information.
In order to test the validity of our measure, we prove it on four Amazons' benchmark instances Amazons', providing some interesting insights for one of them in particular.

As a matter of fact, the definition of such measure could allow platforms to implement a twofold strategy. Detecting whether a node is over-performing or under-performing compared to the average degree level of its neighbours, it is possible -- focusing on over-performing nodes -- to offer them more precise recommendations thanks to to a more accurate inference based on their neighbourhood quality. In turns, in relation to under-performing nodes, it will be possible to exploit the recommendation systems in order to offer them goods or services so that such nodes come into contact with higher quality ones, and prompting higher quality recommendations in the future.

The limitations of such approach are mainly connected to the lack -- in the present literature -- of benchmark instances for empirical tests capable to demonstrate the previous mentioned hypotheses. However, preliminary results obtained provide a positive answer to our research question, and the emerged data properties, together with their envisioned impact in terms of the relevant economic phenomena, points to the suitability of conducting further research on the matter.


\bibliographystyle{spbasic}      

\bibliography{paper}   

\begin{thebibliography}{83}
\providecommand{\natexlab}[1]{#1}
\providecommand{\url}[1]{{#1}}
\providecommand{\urlprefix}{URL }
\expandafter\ifx\csname urlstyle\endcsname\relax
  \providecommand{\doi}[1]{DOI~\discretionary{}{}{}#1}\else
  \providecommand{\doi}{DOI~\discretionary{}{}{}\begingroup
  \urlstyle{rm}\Url}\fi
\providecommand{\eprint}[2][]{\url{#2}}

\bibitem[{Acquisti(2008)}]{acquisti2008identity}
Acquisti A (2008) Identity management, privacy, and price discrimination. IEEE
  Security \& Privacy 6(2):46--50

\bibitem[{Acquisti et~al.(2016)Acquisti, Taylor, and Wagman}]{Acquisti2016}
Acquisti A, Taylor C, Wagman L (2016) {The Economics of Privacy}. Journal of
  Economic Literature 54(2):442--492

\bibitem[{Akerlof(1970)}]{Akerlof1970}
Akerlof GA (1970) {The Market for ''Lemons'' Quality Uncertainty and the Market
  Mechanism}. The Quarterly Journal of Economics 84(3):488--500

\bibitem[{Arpetti(2018)}]{Arpetti2018}
Arpetti J (2018) {Economia della privacy: una rassegna della letteratura (\it
  in italian)}. Rivista di diritto dei media 2:267--297

\bibitem[{Arrow(1958)}]{arrow1958utilities}
Arrow KJ (1958) Utilities, attitudes, choices: A review note. Econometrica:
  Journal of the Econometric Society pp 1--23

\bibitem[{Bakshy et~al.(2012)Bakshy, Rosenn, Marlow, and
  Adamic}]{bakshy2012role}
Bakshy E, Rosenn I, Marlow C, Adamic L (2012) The role of social networks in
  information diffusion. In: Proceedings of the 21st international conference
  on World Wide Web, ACM, pp 519--528

\bibitem[{Barab{\'a}si(2013)}]{barabasi2013network}
Barab{\'a}si AL (2013) Network science. Philosophical Transactions of the Royal
  Society A: Mathematical, Physical and Engineering Sciences 371(1987):20120375

\bibitem[{Barab{\'a}si(2016)}]{barabasi2016network}
Barab{\'a}si AL (2016) Network science. Cambridge university press

\bibitem[{Birke(2013)}]{birke2013social}
Birke D (2013) Social networks and their economics: Influencing consumer
  choice. John Wiley \& Sons

\bibitem[{Bollob{\'a}s(2013)}]{bollobas2013modern}
Bollob{\'a}s B (2013) Modern graph theory, vol 184. Springer Science \&
  Business Media

\bibitem[{Breese et~al.(2013)Breese, Heckerman, and Kadie}]{Breese2013}
Breese JS, Heckerman D, Kadie C (2013) {Empirical Analysis of Predictive
  Algorithms for Collaborative Filtering}. Tech. rep., Microsoft Research

\bibitem[{Briscoe et~al.(2006)Briscoe, Odlyzko, and Tilly}]{BOT2006}
Briscoe B, Odlyzko A, Tilly B (2006) Metcalfe's law is wrong. IEEE Spectrum
  43(7):34--39

\bibitem[{Cabral(2000)}]{Cabral2000}
Cabral LMB (2000) {Introduction to industrial organization}. MIT Press

\bibitem[{Castillejo et~al.(2012)Castillejo, Almeida, and L{\'o}pez-de
  Ipina}]{castillejo2012social}
Castillejo E, Almeida A, L{\'o}pez-de Ipina D (2012) Social network analysis
  applied to recommendation systems: alleviating the cold-user problem. In:
  International Conference on Ubiquitous Computing and Ambient Intelligence,
  Springer, pp 306--313

\bibitem[{Catanzaro et~al.(2005)Catanzaro, Bogu\~n\'a, and
  Pastor-Satorras}]{catanzaro2005generation}
Catanzaro M, Bogu\~n\'a M, Pastor-Satorras R (2005) Generation of uncorrelated
  random scale-free networks. Physical review E 71(2):027103

\bibitem[{Cerqueti et~al.(2018{\natexlab{a}})Cerqueti, Ferraro, and
  Iovanella}]{cerqueti2018new}
Cerqueti R, Ferraro G, Iovanella A (2018{\natexlab{a}}) A new measure for
  community structures through indirect social connections. Expert Systems with
  Applications 114:196--209

\bibitem[{Cerqueti et~al.(2018{\natexlab{b}})Cerqueti, Rotundo, and
  Ausloos}]{cerqueti2018investigating}
Cerqueti R, Rotundo G, Ausloos M (2018{\natexlab{b}}) Investigating the
  configurations in cross-shareholding: a joint copula-entropy approach.
  Entropy 20(2):134

\bibitem[{{Competition and Markets Authority -
  CMA}(2015)}]{CMA-CompetitionandMarketsAuthority2015}
{Competition and Markets Authority - CMA} (2015) The commercial use of consumer
  data report on the cma's call for information. Tech. rep., Competiotion and
  Markets Authority,UK

\bibitem[{{Council of Economic Advisers -
  CEA}(2015)}]{CouncilofEconomicAdvisers-CEA2015}
{Council of Economic Advisers - CEA} (2015) {Big Data and Differential
  Pricing}. Tech. rep., Council of Economic Advisers (CEA) - Executive Office
  of the President of the United States

\bibitem[{Csardi et~al.(2006)Csardi, Nepusz et~al.}]{csardi2006igraph}
Csardi G, Nepusz T, et~al. (2006) The igraph software package for complex
  network research. InterJournal, Complex Systems 1695(5):1--9

\bibitem[{D'Agostino et~al.(2012)D'Agostino, Scala, Zlati{\'c}, and
  Caldarelli}]{d2012robustness}
D'Agostino G, Scala A, Zlati{\'c} V, Caldarelli G (2012) Robustness and
  assortativity for diffusion-like processes in scale-free networks. EPL
  (Europhysics Letters) 97(6):68006

\bibitem[{Erd\H{o}s and Gallai(1960)}]{EG1960}
Erd\H{o}s P, Gallai T (1960) Graphs with prescribed degrees of vertices (in
  hungarian). Matematikai Lapok 11:265--274

\bibitem[{Ezrachi and Stucke(2016{\natexlab{a}})}]{Ezrachi2016a}
Ezrachi A, Stucke ME (2016{\natexlab{a}}) {The rise of behavioural
  discrimination}. European Competition Law Review, ECLR 37(12):485--492

\bibitem[{Ezrachi and Stucke(2016{\natexlab{b}})}]{Ezrachi2016}
Ezrachi A, Stucke ME (2016{\natexlab{b}}) {Virtual competition: the promise and
  perils of the algorithm-driven economy}. Harvard University Press

\bibitem[{Feld(1991)}]{feld1991your}
Feld SL (1991) Why your friends have more friends than you do. American Journal
  of Sociology 96(6):1464--1477

\bibitem[{Firdaus and Uddin(2015)}]{firdaus2015survey}
Firdaus S, Uddin MA (2015) A survey on clustering algorithms and complexity
  analysis. International Journal of Computer Science Issues 12(2):62--85

\bibitem[{Fuller(2018)}]{fuller2018privacy}
Fuller CS (2018) Privacy law as price control. European Journal of Law and
  Economics 45(2):225--250

\bibitem[{Fuller(2019)}]{Fuller}
Fuller CS (2019) Is the market for digital privacy a failure? Public Choice
  180(3-4):353--381

\bibitem[{Galati et~al.(2019)Galati, Bigliardi, Petroni, Petroni, and
  Ferraro}]{galati2019framework}
Galati F, Bigliardi B, Petroni A, Petroni G, Ferraro G (2019) A framework for
  avoiding knowledge leakage: evidence from engineering to order firms.
  Knowledge Management Research \& Practice 17(3):340--352

\bibitem[{Gertz(2002)}]{Gertz2002}
Gertz JD (2002) The purloined personality: Consumer profiling in financial
  services. San Diego L Rev 39:943

\bibitem[{Gilder(1993)}]{gilder1993metcalfe}
Gilder G (1993) Metcalfe's law and legacy. Forbes ASAP 13:1993

\bibitem[{Hakimi(1962)}]{hakimi1962realizability}
Hakimi SL (1962) On realizability of a set of integers as degrees of the
  vertices of a linear graph. Journal of the Society for Industrial and Applied
  Mathematics 10(3):496--506

\bibitem[{Hannak et~al.(2014)Hannak, Soeller, Lazer, Mislove, and
  Wilson}]{Hannak2014}
Hannak A, Soeller G, Lazer D, Mislove A, Wilson C (2014) {Measuring Price
  Discrimination and Steering on E-commerce Web Sites}. In: Proceedings of the
  2014 Conference on Internet Measurement Conference - IMC '14, ACM Press, New
  York, New York, USA, pp 305--318

\bibitem[{Jentzsch(2017)}]{jentzsch2017secondary}
Jentzsch N (2017) Secondary use of personal data: a welfare analysis. European
  Journal of Law and Economics 44(1):165--192

\bibitem[{Kahneman and Tversky(1986)}]{kahneman1986rational}
Kahneman D, Tversky A (1986) Rational choice and the framing of decisions.
  Journal of Business 59(4):251--278

\bibitem[{Kamishima and Akaho(2011)}]{Kamishima2011}
Kamishima T, Akaho S (2011) {Personalized pricing recommender system}. In:
  Proceedings of the 2nd International Workshop on Information Heterogeneity
  and Fusion in Recommender Systems - HetRec '11, ACM Press, New York, New
  York, USA, pp 57--64

\bibitem[{Katarya and Verma(2016)}]{katarya2016collaborative}
Katarya R, Verma OP (2016) A collaborative recommender system enhanced with
  particle swarm optimization technique. Multimedia Tools and Applications
  75(15):9225--9239

\bibitem[{Konstas et~al.(2009)Konstas, Stathopoulos, and
  Jose}]{konstas2009social}
Konstas I, Stathopoulos V, Jose JM (2009) On social networks and collaborative
  recommendation. In: Proceedings of the 32nd international ACM SIGIR
  conference on Research and development in information retrieval, ACM, pp
  195--202

\bibitem[{Kr{\"a}mer and Kalka(2017)}]{kramer2017digital}
Kr{\"a}mer A, Kalka R (2017) How digital disruption changes pricing strategies
  and price models. In: Phantom Ex Machina, Springer, pp 87--103

\bibitem[{Kshetri(2014)}]{kshetri2014big}
Kshetri N (2014) Big data's impact on privacy, security and consumer welfare.
  Telecommunications Policy 38(11):1134--1145

\bibitem[{Lam and Goeksel(2010)}]{lam2010system}
Lam CP, Goeksel M (2010) System and method for utilizing social networks for
  collaborative filtering. US Patent 7,689,452

\bibitem[{Leskovec et~al.(2007)Leskovec, Adamic, and
  Huberman}]{leskovec2007dynamics}
Leskovec J, Adamic LA, Huberman BA (2007) The dynamics of viral marketing. ACM
  Transactions on the Web (TWEB) 1(1):1--39

\bibitem[{Levin(2011)}]{Levin2011}
Levin J (2011) The economics of internet markets. Tech. rep., National Bureau
  of Economic Research, Cambridge, MA

\bibitem[{Linden et~al.(2003)Linden, Smith, and York}]{Linden2003}
Linden G, Smith B, York J (2003) {Amazon.com recommendations: item-to-item
  collaborative filtering}. IEEE Internet Computing 7(1):76--80

\bibitem[{Liu and Lee(2010)}]{liu2010use}
Liu F, Lee HJ (2010) Use of social network information to enhance collaborative
  filtering performance. Expert systems with applications 37(7):4772--4778

\bibitem[{Lu et~al.(2015)Lu, Wu, Mao, Wang, and Zhang}]{lu2015recommender}
Lu J, Wu D, Mao M, Wang W, Zhang G (2015) Recommender system application
  developments: a survey. Decision Support Systems 74:12--32

\bibitem[{L{\"u} et~al.(2012)L{\"u}, Medo, Yeung, Zhang, Zhang, and
  Zhou}]{lu2012recommender}
L{\"u} L, Medo M, Yeung CH, Zhang YC, Zhang ZK, Zhou T (2012) Recommender
  systems. Physics reports 519(1):1--49

\bibitem[{Madureira et~al.(2013)Madureira, den Hartog, Bouwman, and
  Baken}]{madureira2013empirical}
Madureira A, den Hartog F, Bouwman H, Baken N (2013) Empirical validation of
  metcalfe's law: How internet usage patterns have changed over time.
  Information Economics and Policy 25(4):246--256

\bibitem[{Mattioli(2012)}]{mattioli2012orbitz}
Mattioli D (2012) On orbitz, mac users steered to pricier hotels. Wall Street
  Journal 23:2012

\bibitem[{Mavlanova et~al.(2012)Mavlanova, Benbunan-Fich, and
  Koufaris}]{mavlanova2012signaling}
Mavlanova T, Benbunan-Fich R, Koufaris M (2012) Signaling theory and
  information asymmetry in online commerce. Information \& Management
  49(5):240--247

\bibitem[{Metcalfe(2013)}]{metcalfe2013metcalfe}
Metcalfe B (2013) Metcalfe's law after 40 years of ethernet. Computer
  46(12):26--31

\bibitem[{Mikians et~al.(2012)Mikians, Gyarmati, Erramilli, and
  Laoutaris}]{mikians2012detecting}
Mikians J, Gyarmati L, Erramilli V, Laoutaris N (2012) Detecting price and
  search discrimination on the internet. In: Proceedings of the 11th ACM
  Workshop on Hot Topics in Networks, acm, pp 79--84

\bibitem[{Mobasher et~al.(2001)Mobasher, Dai, Luo, and
  Nakagawa}]{mobasher2001improving}
Mobasher B, Dai H, Luo T, Nakagawa M (2001) Improving the effectiveness of
  collaborative filtering on anonymous web usage data. In: Proceedings of the
  IJCAI 2001 Workshop on Intelligent Techniques for Web Personalization
  (ITWP01), pp 53--61

\bibitem[{Newman(2018)}]{newman2018networks}
Newman M (2018) Networks. Oxford university press

\bibitem[{Newman(2002)}]{newman2002assortative}
Newman ME (2002) Assortative mixing in networks. Physical review letters
  89(20):208701

\bibitem[{Newman(2003)}]{newman2003structure}
Newman ME (2003) The structure and function of complex networks. SIAM review
  45(2):167--256

\bibitem[{Nguyen et~al.(2007)Nguyen, Denos, and Berrut}]{nguyen2007improving}
Nguyen AT, Denos N, Berrut C (2007) Improving new user recommendations with
  rule-based induction on cold user data. In: Proceedings of the 2007 ACM
  conference on Recommender systems, ACM, pp 121--128

\bibitem[{Pagallo(2014)}]{Pagallo2014}
Pagallo U (2014) Il diritto nell'et{\`a} dell'informazione: il riposizionamento
  tecnologico degli ordinamenti giuridici tra complessit{\`a} sociale, lotta
  per il potere e tutela dei diritti (in Italian). G. Giappichelli

\bibitem[{Page et~al.(1999)Page, Brin, Motwani, and
  Winograd}]{page1999pagerank}
Page L, Brin S, Motwani R, Winograd T (1999) The pagerank citation ranking:
  Bringing order to the web. Tech. rep., Stanford InfoLab

\bibitem[{Pastor-Satorras et~al.(2001)Pastor-Satorras, V{\'a}zquez, and
  Vespignani}]{pastor2001dynamical}
Pastor-Satorras R, V{\'a}zquez A, Vespignani A (2001) Dynamical and correlation
  properties of the internet. Physical review letters 87(25):258701

\bibitem[{Paterek(2007)}]{paterek2007improving}
Paterek A (2007) Improving regularized singular value decomposition for
  collaborative filtering. In: Proceedings of KDD cup and workshop, vol 2007,
  pp 5--8

\bibitem[{Peel et~al.(2017)Peel, Larremore, and Clauset}]{Peele1602548}
Peel L, Larremore DB, Clauset A (2017) The ground truth about metadata and
  community detection in networks. Science Advances 3(5):e1602548

\bibitem[{Reed(1999)}]{reed1999sneaky}
Reed DP (1999) That sneaky exponential---beyond metcalfe's law to the power of
  community building. Context magazine 2(1)

\bibitem[{Regner and Riener(2017)}]{regner2017privacy}
Regner T, Riener G (2017) Privacy is precious: On the attempt to lift anonymity
  on the internet to increase revenue. Journal of Economics \& Management
  Strategy 26(2):318--336

\bibitem[{Resnick and Varian(1997)}]{Resnick1997}
Resnick P, Varian HR (1997) {Recommender systems}. Communications of the ACM
  40(3):56--58

\bibitem[{Rotundo and D'Arcangelis(2014)}]{rotundo2014network}
Rotundo G, D'Arcangelis AM (2014) Network of companies: an analysis of market
  concentration in the italian stock market. Quality \& quantity
  48(4):1893--1910

\bibitem[{Sarwar et~al.(2001)Sarwar, Karypis, Konstan, and
  Riedl}]{sarwar2001item}
Sarwar B, Karypis G, Konstan J, Riedl J (2001) Item-based collaborative
  filtering recommendation algorithms. In: Proceedings of the 10th
  international conference on World Wide Web, ACM, pp 285--295

\bibitem[{Scott and Carrington(2011)}]{scott2011sage}
Scott J, Carrington PJ (2011) The SAGE handbook of social network analysis.
  SAGE publications

\bibitem[{Shiller(2014)}]{shiller2014first}
Shiller BR (2014) First-degree price discrimination using big data. Tech. rep.,
  Brandeis Univerisity

\bibitem[{Shiller(2015)}]{Shiller2015}
Shiller BR (2015) {Approximating Reservation Prices From Broad Consumer
  Tracking}. Department of Economics, Brandeis University

\bibitem[{Simon(1955)}]{herbert1955behavioral}
Simon H (1955) A behavioral model of rational choice. Quarterly Journal of
  Economics 69(1):99--118

\bibitem[{Simon(1990)}]{simon1990bounded}
Simon HA (1990) Bounded rationality. In: Utility and probability, Springer, pp
  15--18

\bibitem[{Swann(2002)}]{swann2002functional}
Swann GP (2002) The functional form of network effects. Information economics
  and policy 14(3):417--429

\bibitem[{Team et~al.(2013)}]{team2013r}
Team RC, et~al. (2013) R: A language and environment for statistical computing.
  Vienna, Austria

\bibitem[{{The Economist}(2010)}]{TheEconomist2010}
{The Economist} (2010) Clicking for gold. how internet companies profit from
  data on the web. The Economist - A Special Report on Managing Information

\bibitem[{Tsai et~al.(2011)Tsai, Egelman, Cranor, and
  Acquisti}]{tsai2011effect}
Tsai JY, Egelman S, Cranor L, Acquisti A (2011) The effect of online privacy
  information on purchasing behavior: An experimental study. Information
  Systems Research 22(2):254--268

\bibitem[{Van~Hove(2016)}]{van2016testing}
Van~Hove L (2016) Testing metcalfe's law: Pitfalls and possibilities.
  Information Economics and Policy 37:67--76

\bibitem[{Wang and Chen(2003)}]{wang2003complex}
Wang XF, Chen G (2003) Complex networks: small-world, scale-free and beyond.
  IEEE circuits and systems magazine 3(1):6--20

\bibitem[{Xu and Wunsch(2005)}]{xu2005survey}
Xu R, Wunsch DC (2005) Survey of clustering algorithms. IEEE Transaction on
  Neural Networks 16(3):645--678

\bibitem[{Xue et~al.(2005)Xue, Lin, Yang, Xi, Zeng, Yu, and
  Chen}]{xue2005scalable}
Xue GR, Lin C, Yang Q, Xi W, Zeng HJ, Yu Y, Chen Z (2005) Scalable
  collaborative filtering using cluster-based smoothing. In: Proceedings of the
  28th annual international ACM SIGIR conference on Research and development in
  information retrieval, ACM, pp 114--121

\bibitem[{Zhang et~al.(2015)Zhang, Liu, and Xu}]{zhang2015tencent}
Zhang XZ, Liu JJ, Xu ZW (2015) Tencent and facebook data validate metcalfe's
  law. Journal of Computer Science and Technology 30(2):246--251

\bibitem[{Zhao et~al.(2016)Zhao, Zhang, Zhang, and Friedman}]{Zhao2016}
Zhao Q, Zhang Y, Zhang Y, Friedman D (2016) Recommendation based on
  multiproduct utility maximization. Tech. rep., WZB Discussion Paper

\bibitem[{Zhou et~al.(2014)Zhou, Duan, and Piramuthu}]{zhou2014social}
Zhou W, Duan W, Piramuthu S (2014) A social network matrix for implicit and
  explicit social network plates. Decision Support Systems 68:89--97

\end{thebibliography}

\end{document}